\documentstyle[12pt,epsf]{article}
\def\be{\begin{equation}}
\def\ee{\end{equation}}
\def\l{\label}

\def\C{$^o\mbox{C}$}

\def\O{\mbox{\rm O}}
\def\H{\mbox{\rm H}}
\def\C{\mbox{\rm C}}

\def\be{\begin{equation}}
\def\ee{\end{equation}}
\begin{document}
\begin{titlepage}
\title{\large\bf A STUDY OF POLYCARBONYL COMPOUNDS IN MAGNEGASES}
\author{\bf A.K. Aringazin$^{1,2}$ and R.M. Santilli$^2$}
\date{\normalsize
{$^1$Institute for Basic Research, Department of
Theoretical Physics,
Eurasian National University, Astana 473021
Kazakstan}\\[0.5cm]
{$^2$Institute for Basic Research, P.O. Box 1577, Palm
Harbor,}\\
{FL 34682, USA}\\
{\tt ibr@gte.net}\\[0.5cm]
{\small June 5, 2000;
Revised October 10, 2001\\
Final version December 9, 2001} }
\maketitle

\abstract{In this paper we study the structure and thermochemical
properties of some new  polycarbonyl compounds, with particular
attention devoted to the study of (CO)$_n$ complexes, which are
expected to be present in magnegases$^{TM}$. The latter are
anomalous gases produced by {\it Hadronic Reactors$^{TM}$ of
molecular type} [2] (Patented and International Patents Pending)
which expose atoms to the extremely intense electronmagnetic
fields existing at atomic distances from electric arcs in such a
way to create a toroidal distribution of the orbitals of
individual atoms, whether isolated or part of a valence bond.
Polarized atoms, dimers and molecules then attract each other via
opposing magnetic polarities resulting into stable clusters which
constitute  a new chemical species called {\it Santilli's
magnecules} [2]. Some of the numerous open problems in the study
of this intriguing new chemical species are pointed out.}

\end{titlepage}

\section{Introduction}

In a recent paper \cite{1}, we overview the new chemical species
of {\it Santilli's magnecules} \cite{2,3} which has been observed
in magnegases$^{TM}$, the new gas produced by Hadronic
Reactors$^{TM}$ of Molecular type, also called PlasmaArcFlow
Reactors$^{TM}$ (Patented and International Patents pending).

One of the hypotheses on the origin of bonds between diatomic
molecules in magnecules is that atomic orbitals are polarized into
a toroidal distribution when under the influence of very strong
external electromagnetic fields as available at atomic distances
in PlasmaArcFlow$^{TM}$ reactors. In this way, individual
polarized atoms attract each other via opposing magnetic
polarities. Therefore, the magnetic polarization and related bond
here considered exist even for diamagnetic molecules such as $H_2$
\cite{2}.

The isochemical approach developed by Santilli and Shillady
\cite{2,3} has been used to study diatomic molecules
\cite{4,5,6,7}, in order to extend the standard quantum chemical
framework, and to achieve better numerical results on their ground
state energies and bond lengths. This Santilli-Shillady
isochemical model of diatomic molecules uses additional
short-range attractive two-parametric Hulten potential
interactions between valence electron pairs which is assumed to be
due to nonlinear, and other effects originating in the deep
overlapping of wavepackets of atomic electrons in their singlet
valence bond at short distances. The attractive Hulten potential
leads to a specific correlation between two electrons called
isoelectronium \cite{2,3}. The isoelectronium correlations may be
responsible for the anomalous magnetic moments of the molecules,
and thus for the specific bonds in magnecules.

Recent results on two-dimensional two-electron quantum tunnel
effects with dissipation applied to diatomic H-H system \cite{8}
support the isoelectronium-like correlation between two electrons,
in the two-dimensional case. Consequently, we expect that
isoelectronium-like correlations between the electrons due to the
isochemical approach and to the tunnel effects can give an
important contribution to the bonds between molecules in
magnecules.

In addition to the main hypothesis of magnecules, it is
instructive to analyze some other possible compounds which may be
present in magnegases that could have a kind of conventional {\sl
polycarbonyl} structure. Since the mass-spectra of
magnegases$^{TM}$ have not been identified as known compounds
among about 130,000 chemical species, we conclude that the
detected high mass species might be of some {\sl unusual} types of
polycarbonyl compounds, which are absent in computer database of
the mass-spectrometer. Noting that most of extensively studied
aldehydes and ketons (hydrocarbonates containing C=O group) are
{\sl liquids} at room temperatures it is quite natural that they
are not present in magnegases$^{TM}$ in a big percentage. So, we
are led to consider those carbonyl compounds which are expected to
be {\sl gases} at room temperatures.

In the present paper, we focus on some polycarbonyl compounds and
their complexes which may be present in magnegas. An important
note is that we do not study the origin of the specific bonds in
magnecules that can be made elsewhere. Instead, we study some
compounds formed by typical bonds. Consequently, our consideration
is an attempt to identify chemical structures of some components
of magnegas within the framework of typical chemical bonds.
Clearly, such a consideration is helpful in identifying real
structures of magnegas, since one can compare properties of the
polycarbonyl compounds to currently available experimental data on
magnegases$^{TM}$. We believe that such a consideration is a
necessary step toward the unraveling of the intriguing features of
magnegas.

Moreover, the structure of the polycarbonyl compounds studied in
this paper can be taken as a basis for the study of more general
magnecules. The fact that one of the suggested structures of
magnecules, CO$\times$CO$\times\cdots\times$CO \cite{2}, where
$\times$ denotes a (magnetic) bond and CO is carbon monoxide, is
known in practical chemistry, carbon monoxide complex (CO)$_6$,
serves us as a strong {\sl experimental} ground to focus on the
polycarbonyl compounds. Another interesting experimental fact is
that some polycarbonyl compounds are known to be {\sl gases}, at
room temperatures, so that they can be present in
magnegases$^{TM}$.

In Sec.~\ref{SomeExamples} we present some examples of
polycarbonyl compounds. In Sec.~\ref{COnComplex}, we study the
structure and combustion of (CO)$_n$ complex. In
Sec.~\ref{HydrogenBridges}, we consider possible hydrogen bonds
between (CO)$_n$  complexes. In Sec.~\ref{CCOn}, we consider some
other possible types of polycarbonyl compounds. In
Sec.~\ref{CObond}, we briefly outline the properties of the
carbonyl C=O bond which are helpful in understanding of the
behavior of polycarbonyl compounds under the influence of external
electromagnetic field.

Our general remark is that the term "polycarbonyl compound" can be
treated in a rather general form, without specifying  the
character of some bonds (conventional, or unconventional), but
stressing only the presence of several C=O groups. Indeed, even
the bond between C and O in the magnecules
CO$\times$CO$\times\cdots\times$CO \cite{2}, could not be of a
conventional type, with some electronic effects playing an
important role with interesting properties.

\section{Some examples of polycarbonyl compounds}\l{SomeExamples}

We start by some characteristic examples of polycarbonyl
compounds, i.e. the compounds containing several carbonyl groups
C=O.

\begin{figure}[ht]
\begin{center}
\unitlength=0.5mm
\begin{picture}(100,80)
\put(17,41){O} \put(39,10){\line(0,0){7}}
\put(41,10){\line(0,0){7}} \put(40,30){\line(0,0){7}}
\put(40,50){\line(0,0){7}} \put(39,70){\line(0,0){7}}
\put(41,70){\line(0,0){7}} \put(37, 1){C} \put(37,21){O}
\put(37,41){Ni} \put(37,61){O} \put(37,81){C} \put(57,41){O}
\put(8,43){\line(3,0){7}} \put( 8,45){\line(3,0){7}}
\put(27,44){\line(3,0){7}} \put(47,44){\line(3,0){7}}
\put(66,43){\line(3,0){7}} \put(66,45){\line(3,0){7}}
\put(0,41){C} \put(75,41){C}
\end{picture}
\end{center}
\caption{Ni(CO)$_4$. Dissociation of this gas, Ni(CO)$_4 \to$ Ni +
4CO, occurs at temperature $T=200^o\C$.}
\label{Ni}
\end{figure}
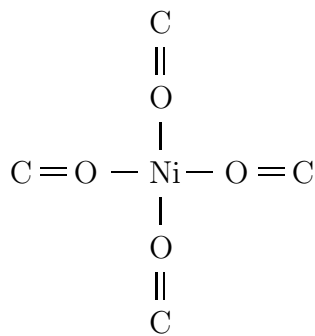

Cobalt hydrocarbonyl, HCo(CO)$_4$, containing both H and CO, is a
gas, at room temperatures. Such type of a compound is known as
somewhat unusual because a neutral metal is bonded to carbon
monoxide CO, which mostly conserves its own properties. Another
example is magnesium carbonyl, (CO)$_5-$Mn$-$Mn$-$(CO)$_5$
(melting point is $66^o\C$), where the bond Mn$-$Mn is about 40
kcal/mol. Also, it is interesting to note that nickel carbonyl,
Ni(CO)$_4$, shown in Fig.~\ref{Ni}, formed from Ni and CO at
$T=80\C$, is a gas at room temperatures, and dissociates,
Ni(CO)$_4 \to$ Ni + 4CO, at $T=200\C$.

Thus, the binding energy of the bond between Ni and each CO is
about 30 kcal/mol, which is within the range given by the
estimation \cite{1},
\be\l{BEmag}
B[magnecule] > 25...30 \mbox{ kcal/mol},
\ee
of the average binding energy between the molecules in magnecule.
In general, various carbonyl compounds can be produced from, e.g.,
some glicoles (containing the group -C-OH).

As mentioned in ref. \cite{1} there is a case in which an
explosive compound, potassium carbonyl, 6CO + 6K$\to$
K$_6$(CO)$_6$, is produced. Such a compound is used to obtain
unusual carbon monoxide (carbon monoxide complex), (CO)$_6$. This
compound is believed to exist due to a polymerization, namely, the
structure (CO)$_6$ is thought as being given by a sequential
joining of separate CO (monomers) to a linear chain of CO
molecules (polymer) owing to the C-C bonds.

\section{(CO)$_n$ complex}\l{COnComplex}

\subsection{Structure}

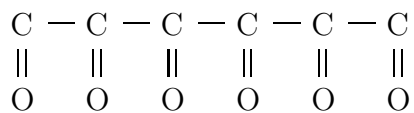
\begin{figure}[ht]
\begin{center}
\unitlength=0.5mm
\begin{picture}(130,50)
\put(19,30){\line(0,0){7}} \put(21,30){\line(0,0){7}}
\put(17,21){O} \put(17,41){C} \put(39,30){\line(0,0){7}}
\put(41,30){\line(0,0){7}} \put(37,21){O} \put(37,41){C}
\put(59,30){\line(0,0){7}} \put(61,30){\line(0,0){7}}
\put(57,21){O} \put(57,41){C} \put(79,30){\line(0,0){7}}
\put(81,30){\line(0,0){7}} \put(77,21){O} \put(77,41){C}
\put(99,30){\line(0,0){7}} \put(101,30){\line(0,0){7}}
\put(97,21){O} \put(97,41){C} \put(119,30){\line(0,0){7}}
\put(121,30){\line(0,0){7}} \put(117,21){O} \put(117,41){C}
\put(27,44){\line(3,0){7}} \put(47,44){\line(3,0){7}}
\put(67,44){\line(3,0){7}} \put(87,44){\line(3,0){7}}
\put(107,44){\line(3,0){7}}
\end{picture}
\end{center}
\vskip -10mm \caption{Possible polymer structure of (CO)$_6$
complexes.}
\label{CO6}
\end{figure}

In Fig.~\ref{CO6} we present a linear chain of CO molecules
(polymer) as a possible structure of (CO)$_6$.

Owing to C-C bonds, a typical polymerization, e.g., of
propylene,\newline CH(CH$_3$)CH$_2$, is characterized by binding
energies of about 73...83 kcal/mol, with the reaction heat of
about $\Delta H=-20$~kcal/mol per each molecule of the linear
chain of polypropylene.

We see that the typical value of the binding energy of C$-$C is
much bigger than 30...35~kcal/mol. However, some other types of
intermolecular interaction between CO can also make a contribution
here because of the specific electronic structure of CO molecule,
and the real structure of (CO)$_6$ may be {\sl different} from
that shown in Fig.~\ref{CO6}. So, we could expect lower values of
the binding energy between CO molecules in (CO)$_6$, recalling
that the above mentioned Ni(CO)$_4$ dissociates at $T=200$C.

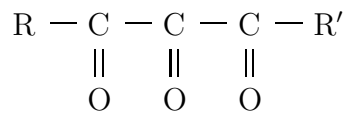
\begin{figure}[ht]
\begin{center}
\unitlength=0.5mm
\begin{picture}(130,50)
\put(17,41){R} \put(39,30){\line(0,0){7}}
\put(41,30){\line(0,0){7}} \put(37,21){O} \put(37,41){C}
\put(59,30){\line(0,0){7}} \put(61,30){\line(0,0){7}}
\put(57,21){O} \put(57,41){C} \put(79,30){\line(0,0){7}}
\put(81,30){\line(0,0){7}} \put(77,21){O} \put(77,41){C}
\put(97,41){R$^\prime$}
\put(27,44){\line(3,0){7}} \put(47,44){\line(3,0){7}}
\put(67,44){\line(3,0){7}} \put(87,44){\line(3,0){7}}
\end{picture}
\end{center}
\vskip -10mm \caption{A view of Tricarbonyl compounds.}
\label{Tri}
\end{figure}

Also, it is known that in tricarbonyl compounds (see
Fig.~\ref{Tri}), the central carbonyl group C=O is highly
reactional since it is weakly bonded to the two neighbor C=O
groups, and it can be easily decarbonylized (releasing of the
central C=O group as carbon monoxide gas) by using catalysis with,
e.g., AlCl$_3$. For the same reason the central carbonyl group C=O
in tricarbonyl compounds easily reacts with water, becoming the
HO-C-OH group due to the reaction C=O + H$_2$O $\to$ HO-C-OH.

It is important to note that there may be also {\sl cyclic}
polycarbonyl structures (with all C atoms single bonded to each
other to form a circle) which are characterized by higher
stability than the linear ones so they could be either gas or
liquid, at room temperatures.

\subsection{Combustion}

We should note that the bond C=O in (CO)$_6$ evidently is weaker
than its counterpart in a single carbon monoxide C=O. Moreover,
some energy is related to a dissociation of (CO)$_6$ to 6CO.

In general, this could lead to a different value of the combustion
reaction heat of (CO)$_6$ per CO molecule. Namely, the reaction
\be
(\C\O)_6 + 3\O_2 \to 6\C\O_2
\ee
could give a different value of the reaction heat $\Delta H$ than
that of the reaction
\be
6\C\O + 3\O_2 \to 6\C\O_2.
\ee

\begin{table}
\begin{center}
\begin{tabular}{|cc||cc|}
\hline
Diatomic molecules &&Diatomic molecules &\\
\hline
H--H              &104.2 & C=O & 255.8\\
\hline
O=O              &119.1 & $N\equiv N$ &225.8\\
\hline \hline
Manyatomic molecules &&Manyatomic molecules &\\
\hline
C--O                 &85.5    & O--H & 110.6\\
\hline
C=O in CO$_2$       &192.0   & O--O & 35\\
\hline
C=O in formaldegide &166     & C--H & 98.7\\
\hline
C=O in aldehydes    &176     & C--C & 82.6\\
\hline
C=O in ketons       &179     & C=C & 145.8 \\
\hline
C=N                 &147     & C$\equiv$C &199.6\\
\hline
\end{tabular}
\caption{Average binding energies, kcal/mol [9]. $T=25\C$.}
\l{Table1}
\end{center}
\end{table}

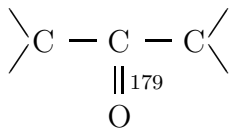
\begin{figure}[ht]
\begin{center}
\unitlength=0.5mm
\begin{picture}(70,50)
\put(43,31){\scriptsize 179}  
\put(17,41){C} \put(39,30){\line(0,0){7}}
\put(41,30){\line(0,0){7}} \put(37,21){O} \put(37,41){C}
\put(57,41){C}
\put(16,43){\line(-2,-3){5}} 
\put(16,45){\line(-2,3){5}} \put(27,44){\line(3,0){7}}
\put(47,44){\line(3,0){7}}
\put(64,45){\line(2,3){5}}  
\put(64,43){\line(2,-3){5}}
\end{picture}
\end{center}
\vskip -10mm \caption{Carbonyl group C=O in ketons.} \label{keton}
\end{figure}

Below, we make a crude estimate by taking the average value of the
binding energy of C=O bond in Fig.~\ref{CO6} equal to that in
ketons, 179 kcal/mol (see Table~1), since most ketons are
characterized by carbonyl group bonded to two C atoms as
schematically shown in Fig.~\ref{keton}.

We assume also that each C-C bond in Fig.~\ref{CO6} requires 82.6
kcal/mol (see Table~1). There are five bonds C-C in Fig.~\ref{CO6}
so we have 5$\times$82.6=413 kcal/mol. There are six C=O bonds in
Fig.~\ref{CO6} so we have 6$\times$179 = 1074 kcal/mol. On the
other hand, six separate $C=O$ molecules have in total
6$\times$255.8 = 1534.8 kcal/mol due to their bonds. Hence, the
balance is 415+1074-1534.8=-47.8 kcal/mol, and the dissociation
\be\l{dCO6}
(\C\O)_6 \to 6\C\O + 47.8 \mbox{ kcal/mol}
\ee
appears to be an exothermic reaction, with about $47.8/6 \simeq 8$
kcal/mol per each CO. This conservative estimate shows that carbon
monoxide gas consisting of (CO)$_6$ complexes might give bigger
energy release than that of the carbon monoxide gas consisting of
separate CO molecules. Namely, instead of the reaction
\be\l{COOCO}
\C\O + \O_2/2 \to \C\O_2 + 68.7 \mbox{ kcal/mol},
\ee
for the case of (CO)$_6$, we consider the sequence
\begin{eqnarray}
(\C\O)_6 \to 6\C\O + 47.8 \mbox{ kcal/mol}, \\
\C\O + \O_2/2 \to \C\O_2 + 68.7 \mbox{ kcal/mol},
\end{eqnarray}
which effectively could give $68.7+8 = 76.8$ kcal/mol per CO,
instead of 68.7 kcal/mol, i.e. about 12\% bigger heat released.

It is a consequence of the existence of (CO)$_6$ that there might
exist lower and higher mass carbon monoxide complexes, (CO)$_n$,
with $n=2,3,4, \dots$ For the general case of dissociation of
conceivable (CO)$_n$ complexes of linear type depicted in
Fig.~\ref{CO6},
\be\l{gdCOn}
(\C\O)_n \to n\C\O,
\ee
the dissociation energy per produced CO molecule thus is
\be\l{bal}
\Delta H=\frac{1}{n}\left(82.6(n-1)+179n-255.8n\right) =
          5.8-\frac{82.6}{n} \mbox{ kcal/mol}.
\ee
It is interesting to note that this energy is about zero for
$n=14$, i.e. there appears to be no considerable difference in
calculated combustion heat between the conceivable (CO)$_{14}$ gas
and usual CO gas.

The above estimate can be formulated in terms of the binding
energies $B$[C-C] and $B$[C=O], to get a general expression of the
average dissociation energy of linear (CO)$_n$ ($n \geq 2$) per
each CO,
\be\l{balg}
\Delta H=\frac{1}{n}(B[\C-\C](n-1)+B[\C=\O]n-255.8n)
         \mbox{ kcal/mol},
\ee
where presently unknown precise values of $B$[C-C] and $B$[C=O] in
the linear (CO)$_n$ complex can be evaluated from theory or
experiment.

In fact, in the presence of one $C=O$ group the above C-C bond is
known to be much weaker than the above used average value 82.6
kcal/mol, namely, $B$[C-C] = 73 kcal/mol. So the estimations
(\ref{dCO6}) and (\ref{bal}) must be corrected. Inserting $B$[C-C]
= 73 kcal/mol and $B$[C=O] = 179 kcal/mol into (\ref{balg}), we
get the average dissociation energy of (CO)$_n$ ($n \geq 2$) per
each CO,
\be\l{bal2}
\Delta H=\frac{1}{n}(73(n-1)+179n-255.8n) =
         -3.8-\frac{73}{n} \mbox{ kcal/mol}.
\ee
This estimate shows that dissociation of (CO)$_n$ complex
(\ref{gdCOn}) is {\sl always} exothermic (at any value of $n\geq
2$).

Returning to our example (CO)$_6$, the dissociation energy
(\ref{bal2}) gives the value 96/6 = 16 kcal/mol per each CO
produced. The reaction (\ref{dCO6}) now reads
\be\l{2dCO6}
(\C\O)_6 \to 6\C\O + 96 \mbox{ kcal/mol},
\ee
and appears to be highly exothermic, with about 16 kcal released
per each CO. Therefore, in contrast to the combustion
(\ref{COOCO}), this effectively yields 68.7+16=84.7 kcal/mol, as
compared to 68.7 kcal/mol, {\sl i.e. about 23\% bigger heat
released}, due to the dissociation of (CO)$_6$ to $6CO$.

In the general case of combustion of (CO)$_n$, we get
\be\l{COnincr}
\frac{3.8+72/n}{68.7}100 \ \%
\ee
increase of the energy release per CO, as compared to the usual CO
gas.

On the other hand, we see that one mole of (CO)$_6$ gives 6 moles
of CO, thus the reaction entropy $\Delta S$ is expected to be of
high value. As we mentioned in ref. \cite{1}, for carbon monoxide
CO the reaction constant $K$ of its formation increases at higher
temperatures (it is harder to dissociate CO at high temperatures
than at low temperatures) because of high value of the reaction
entropy. Similarly, we might conclude that it is harder to
dissociate (CO)$_6$ at high temperatures than at low temperatures.
Thus, (CO)$_n$ complexes could survive high temperatures, being
metastable at room temperatures. We expect that the dissociation,
(CO)$_n \to$ $n$CO occurs most effectively only at some small
fixed temperature range depending on $n$.

In the general case, for one mole of (CO)$_n$ gas, we have the
sequence of reactions,
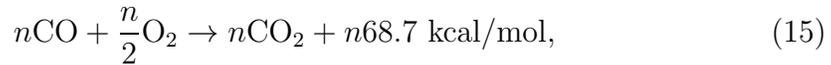
\begin{eqnarray}\l{dCOn}
(\C\O)_n \to n\C\O + n3.8+73 \mbox{ kcal/mol}, \\
n\C\O + \frac{n}{2}\O_2 \to n\C\O_2 + n68.7 \mbox{ kcal/mol},
\end{eqnarray}
so that the total heat released is
\be
\Delta H = -n72.5 - 73 \simeq -(n+1)73 \mbox{ kcal/mol}.
\ee

We see that the combustion of one mole of (CO)$_n$ gas gives {\sl
much bigger} heat, $\Delta H =-511$ kcal for the case of (CO)$_6$
gas, than one mole of CO gas characterized by $\Delta H =-68.7$
kcal/mol, {\sl i.e. 511/68.7= 7.4 times bigger heat release per
mole}. This is not only a mere consequence of the fact that one
mole of (CO)$_n$ dissociates to $n$ moles of CO but also due to
the dissociation heat presented in Eq. (\ref{dCOn}).

In any case, the above carbon monoxide complex, (CO)$_6$, is a
direct confirmation that the conjectured CO$\times$CO bond
\cite{2} really exists, as it is known in practical chemistry. So
the complex
\be\l{COn}
\C\O\times \C\O\times\cdots\times \C\O\times \C\O,
\ee
where "$\times$" denotes a bond, is a good candidate of a
magnecule.

\section{Hydrogen bridges}\l{HydrogenBridges}

The complex of type (\ref{COn}) would give a mass-spectrum which
exhibits a periodicity in molecular masses. Indeed, the weakest
bonds in (\ref{COn}) are evident (namely, those denoted by
$\times$) so that the complex (\ref{COn}), under the influence of
electronic beam in mass-spectrometer, should dissociate to an
integer number of (ionized) CO molecules. However, the
mass-spectra of magnegases does not reveal periodicity 28 a.m.u.
as the smallest step. Instead, we observe almost randomly
distributed masses of the charged fragments, with the minimal mass
difference being 1 a.m.u. Thus, we are led to the assumption that
there are some other types of magnecules in magnegases, in
addition to (\ref{COn}). We expect the presence of H atoms in
magnecules which could provide (multiple) hydrogen bonds.

\begin{figure}[ht]
\begin{center}
\unitlength=0.5mm
\begin{picture}(130,100)
\put(19,90){\line(0,0){7}} \put(21,90){\line(0,0){7}}
\put(17,81){O} \put(17,101){C} \put(39,90){\line(0,0){7}}
\put(41,90){\line(0,0){7}} \put(37,81){O} \put(37,101){C}
\put(59,90){\line(0,0){7}} \put(61,90){\line(0,0){7}}
\put(57,81){O} \put(57,101){C} \put(79,90){\line(0,0){7}}
\put(81,90){\line(0,0){7}} \put(77,81){O} \put(77,101){C}
\put(99,90){\line(0,0){7}} \put(101,90){\line(0,0){7}}
\put(97,81){O} \put(97,101){C} \put(119,90){\line(0,0){7}}
\put(121,90){\line(0,0){7}} \put(117,81){O} \put(117,101){C}
\put(27,104){\line(3,0){7}}  
\put(47,104){\line(3,0){7}} \put(67,104){\line(3,0){7}}
\put(87,104){\line(3,0){7}} \put(107,104){\line(3,0){7}}
\put(40,50){\line(0,0){7}}  
\put(37,60){H} \put(40,69){\line(0,0){2}}
\put(40,72){\line(0,0){2}} \put(40,75){\line(0,0){2}}
\put(80,50){\line(0,0){7}}  
\put(77,60){H} \put(80,69){\line(0,0){2}}
\put(80,72){\line(0,0){2}} \put(80,75){\line(0,0){2}}
\put(5,41){$\delta\oplus$}  
\put(5,21){$\delta\ominus$} 
\put(5,81){$\delta\ominus$} 
\put(5,101){$\delta\oplus$} 
\put(19,30){\line(0,0){7}} \put(21,30){\line(0,0){7}}
\put(17,21){O} \put(17,41){C} \put(39,30){\line(0,0){7}}
\put(41,30){\line(0,0){7}} \put(37,21){O} \put(37,41){C}
\put(59,30){\line(0,0){7}} \put(61,30){\line(0,0){7}}
\put(57,21){O} \put(57,41){C} \put(79,30){\line(0,0){7}}
\put(81,30){\line(0,0){7}} \put(77,21){O} \put(77,41){C}
\put(99,30){\line(0,0){7}} \put(101,30){\line(0,0){7}}
\put(97,21){O} \put(97,41){C} \put(119,30){\line(0,0){7}}
\put(121,30){\line(0,0){7}} \put(117,21){O} \put(117,41){C}
\put(27,44){\line(3,0){7}} 
\put(47,44){\line(3,0){7}} \put(67,44){\line(3,0){7}}
\put(87,44){\line(3,0){7}} \put(107,44){\line(3,0){7}}
\end{picture}
\end{center}
\vskip -10mm \caption{Conjectured hydrogen bonds between two
linear (CO)$_6$.}
\label{hyd}
\end{figure}
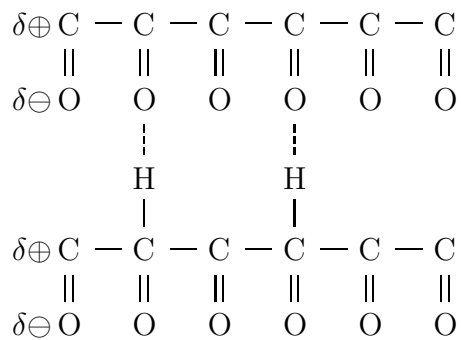

As a possibility of hydrogen binding between polycarbonyl
compounds, we conjecture a double hydrogen bond between two linear
(CO)$_6$ as shown in Fig.~\ref{hyd}.

This association could survive room temperatures, and can exist in
a gas state since each of the hydrogen bond provides about 5...10
kcal/mol, leading to metastability of the compound. Here, we use
the fact that the carbonyl C=O bond in ketons is highly polarized,
as it is shown in Fig.~\ref{Polar}.

It is remarkable to note that such an association would have total
dipole moment as a vector sum of two big dipole moments;
schematically, some three C=O bonds (e.g., 1, 3, and 5) in each
(CO)$_6$ are directed up, and the other three (e.g., 2, 4, and 6)
are directed down. Thus, it would attract more number of H atoms
(by induced polarization) and CO molecules which are close to one
of these two big dipoles. Also, this association would reveal a
big adhesion to some solid surfaces by inducing a polarization of
the surface molecules. Linear growth of this association is
terminated by heat excitations and collisions of the complexes so
bigger number of high mass structures are expected to appear
automatically at lower temperatures.

More detailed studies on the geometry and total dipole moments of
the (CO)$_6$ complex, and its possible associations via hydrogen
bridges, are needed to make accurate estimates. It should be noted
that C-C bonds in (CO)$_6$ are mostly symmetric so that they give
low intensity infrared peaks. Raman spectrum would show presence
of such symmetric bonds.

On the other hand, experimental measurements of the dielectric
constant of magnegases$^{TM}$ could give us valuable information
on the averaged total polarization of magnegas components (see Eq.
(\ref{Pdiel}) below).

In general, it is natural to expect various combinations of
hydrogen bonds between several (CO)$_n$ complexes. The example
shown in Fig.~\ref{hyd} illustrates only one possibility. Another
simple example is given by two carbon monoxide CO molecules bonded
by hydrogen bridge,
\be
\C=\O\cdots \H\!\! \mbox{ --- } \C=\O,
\ee
or two carbon monoxide $(CO)_2$ complexes bonded to each other by
hydrogen bridge, and several H atoms (weakly) bonded to (CO)$_n$
complex at O atoms.

We note that the presence of hydrogen bonds could be detected in a
gas state NMR spectral analysis of magnegases. We note also that
the usual contribution of hydrogen bonds makes some infrared peaks
wider. The higher combustion energy release of the association
shown in Fig.~\ref{hyd} could be also due to an internal tension
of this structure.

\section{C(CO)$_n$ complex}\l{CCOn}

\begin{figure}[ht]
\begin{center}
\unitlength=0.5mm
\begin{picture}(100,80)
\put(17,41){C} \put(39,10){\line(0,0){7}}
\put(41,10){\line(0,0){7}} \put(40,30){\line(0,0){7}}
\put(40,50){\line(0,0){7}} \put(39,70){\line(0,0){7}}
\put(41,70){\line(0,0){7}} \put(37, 1){O} \put(37,21){C}
\put(37,41){C} \put(37,61){C} \put(37,81){O} \put(57,41){C}
\put(8,43){\line(3,0){7}} \put( 8,45){\line(3,0){7}}
\put(27,44){\line(3,0){7}} \put(47,44){\line(3,0){7}}
\put(66,43){\line(3,0){7}} \put(66,45){\line(3,0){7}}
\put(0,41){O} \put(75,41){O}
\end{picture}
\end{center}
\caption{A possible compound $C(CO)_4$.}
\label{CCO4}
\end{figure}
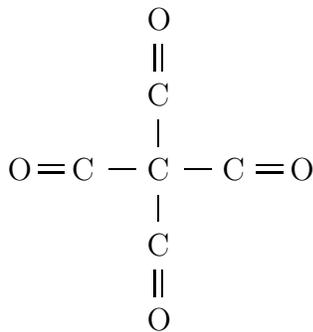

Another possible type of polycarbonyl compound of interest is the
following. In Ni(CO)$_4$ shown in Fig.~\ref{Ni}, we can replace
the Ni atom with C atom, convert the C=O bonds and conjecture the
existence of the compound C(CO)$_4$ depicted in Fig.~\ref{CCO4}.
Note that infrared spectra of such a compound would reveal only
the C=O bonds because of the almost symmetric character of all the
C-C bonds. One can extend this compound by noting that the C atoms
in each of four C=O could give {\sl further} bonds with additional
C=O molecules, thus providing more complicated structure,
C(CO)$_n$, having high molecular mass. One can easily estimate the
dissociation heat of C(CO)$_n$ by using the simple technique
applied to (CO)$_n$ in Sec.~\ref{COnComplex}. Namely, it equals
\be\l{dCCO4}
\Delta H = kB[\C-\C]+ nB[\C=\O] - n255.8 \mbox{ kcal/mol},
\ee
where $k$ is number of C-C bonds, and $B$[C=O] and $B$[C-C] are
average binding energy of C=O and C-C bonds in C(CO)$_n$ complex,
respectively. Note that C(CO)$_n$ dissociates to carbon monoxide
CO and pure carbon C.

It is interesting to analyze such a kind of compounds in order to
make accurate estimations of binding energies of the bonds. In
particular, we expect that the energy $B$[C-C] in C(CO)$_4$ is
smaller than $B$[C-C] in (CO)$_6$, hence we would have bigger
combustion energy released per CO.

\begin{figure}[ht]
\begin{center}
\unitlength=0.5mm
\begin{picture}(130,50)
\put(39,30){\line(0,0){7}} \put(41,30){\line(0,0){7}}
\put(37,21){O} \put(37,41){C} \put(57,41){O} \put(77,41){O}
\put(99,30){\line(0,0){7}} \put(101,30){\line(0,0){7}}
\put(97,21){O} \put(97,41){C} \put(27,44){\line(3,0){7}}
\put(47,44){\line(3,0){7}} \put(67,44){\line(3,0){7}}
\put(87,44){\line(3,0){7}} \put(107,44){\line(3,0){7}}
\end{picture}
\end{center}
\vskip -10mm \caption{Possible peroxide group, -O-O-, in a linear
chain of CO monomers.}
\label{OO}
\end{figure}
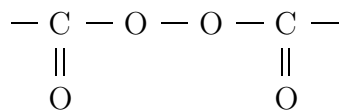

The -O-O- group serving as a peroxide bridge could be present in
(CO)$_n$ and C(CO)$_n$ as well, as shown in Fig.~\ref{OO}.

In accordance with the mostly symmetric character of such bonds,
they give low intensity infrared peaks. Thus, such bonds could not
be observed in infrared spectra \cite{2} of magnegases$^{TM}$.
Again, Raman spectra would be of much help here as they allow to
identify all symmetric bonds in compounds.

\section{C=O bond}\l{CObond}

In order to analyze C=O bond as a group in compounds, we outline
below properties of this bond in ketons. The structures of ketons
is shown in Fig.~\ref{keton}. These properties are also
characteristic of carbon monoxide and some other polycarbonyl
compounds, as a consequence of high strength of the C=O bond, as
compared to the bonds C-C, C=C, C-O, O-H, H-H, and O-O. In fact,
only C$\equiv$C bond has a higher strength (see Table~1).

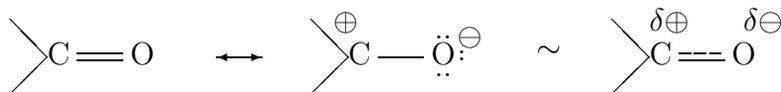
\begin{figure}[ht]
\begin{center}
\unitlength=0.5mm
\begin{picture}(200,30)
\put(10,12){\line(-1,1){10}}  
\put(10,12){\line(-1,-1){10}} 
\put(10,10){C}
\put(18,11){\line(3,0){12}}    
\put(18,13){\line(3,0){12}} \put(32,10){O}
\put(55,12){\vector(3,0){12}}   
\put(67,12){\vector(-3,0){12}}
\put(90,12){\line(-1,1){10}}  
\put(90,12){\line(-1,-1){10}} 
\put(90,10){C}
\put(98,12){\line(3,0){12}}   
\put(112,10){O}
\put(140,12){$\sim$}          
\put(86,19){$\oplus$}          
\put(119,15){$\ominus$}          
\put(113,17){..}          
\put(119,11){.}          
\put(119,13){.}          
\put(113,7){..}          
\put(170,12){\line(-1,1){10}}  
\put(170,12){\line(-1,-1){10}} 
\put(170,10){C}
\put(178,11){\line(3,0){12}}  
\put(178,13){\line(3,0){3}}   
\put(182,13){\line(3,0){3}}   
\put(186,13){\line(3,0){3}}   
\put(192,10){O}
\put(170,19){$\delta\oplus$}          
\put(195,19){$\delta\ominus$}          
\end{picture}
\end{center}
\caption{Bipolar structure of the bond C=O in ketons.}
\label{Polar}
\end{figure}

The bond C=O in ketons is known both as very strong and very
reactive (high reaction rate). It should be noted that the energy
of this bond (179 kcal/mol in ketons) is bigger than the sum of
two single C-O bonds (2$\times$85.5 = 171). This is in contrast
with the double bond $C=C$ (145.8 kcal/mol) which is weaker than a
sum of two single bonds C-C (2$\times$82.6 = 165.2). Also, the
bond C=O in ketons is of about 40...50\% ionic character due to
big resonance bipolar contributions, with oxygen O being charged
negative and carbon C being charged positive, as shown in
Fig.~\ref{Polar}. This is due to different electronegativities of
C and O (2.55 and 3.44, respectively). The infrared spectra of the
carbonyl group in ketons is known to be at 1705...1740 cm$^{-1}$.
Dipole moments of most ketons are about 2.7 D.

The above outline allows us to conclude that carbonyl bond C=O in
carbon monoxide CO, ketons, and polycarbonyl compounds, is highly
polarized and, therefore, this bond and its electron structure are
highly sensitive to an external electromagnetic field. Note also
that four electrons (two of C and two of O), among ten electrons,
are available to provide conventional bonds, and do not contribute
the double bond C=O.

Note also that the constant dipole moment of the carbon monoxide
C=O molecule is $\mu=3.2$ Debay; for C-O bond it is $\mu=1.5$ D,
and for symmetrical H-H, O=O, and C-C bonds it is zero. A
considerable numerical value of $\mu$ for C=O implies that CO
molecules, as highly polarized molecules, tend to have a strong
order in the presence of an external electromagnetic field despite
a heat disordering.

In general, the total polarization of one mole of a gas is the sum
of induced, $P_\alpha$,  and constant, $P_\mu$, polarizations,
\be\l{Ptot}
P =\frac{4\pi N_A}{3}\alpha + \frac{4\pi N_A}{3}\frac{\mu^2}{3kT},
\ee
which can be estimated from the experimental value of the
dielectric constant $\varepsilon$ of a gas,
\be\l{Pdiel}
P= \frac{\varepsilon-1}{\varepsilon+2}\frac{M}{d},
\ee
where $M$ is molecular mass, and $d$ is density of a gas.

We expect the presence of about 2...20 carbonyl groups C=O in a
cluster so the total electric polarization of magnecule can be
roughly estimated by vector sum of individual polarizations of CO
if one identifies structure of the magnecule.

The observed high adhesive property of magnegases$^{TM}$ \cite{2}
could be interpreted due to high value of the electric dipole
moment of the clusters containing several CO molecules acting in
parallel.

\section{Conclusions}

In this paper, an attempt has been made to identify the structure
of magnecules by analyzing the structure and properties of some
polycarbonyl compounds. We stress that we have not tried to
identify the origin of the specific bonds in magnecules.
Therefore, the numerical values of the binding energies used in
this paper are approximate, and should be estimated more precisely
in accord to specific models of magnecules. Nevertheless, we have
shown possible reasons of increased combustion energy content of
some polycarbonyl compounds viewed as candidates to magnecules, as
compared to that of usual carbon monoxide CO and hydrogen H$_2$
gases.  More experimental data and further theoretical study are
needed to identify the structure of magnegases and analyze their
combustion.

It is intriguing to note that, despite measurements of magnegases
conducted over a two years period by using most of available
detection methods and equipment (such as IRD, GC, MS, GC-MS,
GC-MS/IRD, FTIR, and other methods), the true and detailed
chemical composition of magnegas remains wastly unknown at this
writing, thus stimulating new research [2].

In fact, according to conventional infrared spectroscopy,
magnegas$^{TM}$ produced from water as liquid feedstock is
essentially composed of H$_2$ and CO in about equal volume
percentages, plus traces of O$_2$, H$_2$O, and CO$_2$. Its exhaust
from internal combustion engine) is essentially composed of: 50\%
to 60\% of H$_2$O (as water vapor); 10\% to 15\% of O$_2$; 10\% to
15\% of C (estimate); 3\% to 7\% of CO$_2$; the remaining
components being inert atmospheric gases.

These data can be naively interpreted as follows. 50\% to 60\% of
H$_2$O in the exhaust means that there were 50\% to 60\% of 2H in
magnegas; 10\% to 15\% of C in the exhaust means that there was
incomplete combustion of magnegas (otherwise, we would not have
pure carbon in the exhaust), and that C is released from
magnecules; 10\% to 15\% of O$_2$ in the exhaust means, again,
that there was incomplete combustion of magnegas; 3\% to 7\% of
CO$_2$ in the exhaust disproves the expectation according to
conventional quantum chemistry that about 50\% of magnegas from
water is constituted by CO$_2$, because in this case the CO$_2$
percentage in the exhaust should have been of the order of {\sl
ten times bigger}.

By putting all available information together, including the
anomalous energy content of magnegas, and as sated in ref. [2],
available data establish that magnegas produced from water as a
feedstock is indeed composed 50\% H {\sl atoms}, 25\% C {\sl
atoms} and 25\% O {\sl atoms}, although these  atoms are clustered
into magnecules composed of individual and unbounded H, C and O
atoms, radicals OH and CH, single and double bonds C-O, as well as
conventional molecules H$_2$, CO, O$_2$, CO$_2$, H$_2$O, and other
molecules in relative percentages unknown at this time. The
erroneous reading of 50\% CO in magnegas can be readily explained
by the fact that the detecting frequency can trigger the {\sl
creation of} CO {\sl by the instrument}, e.g., via the conversion
of all single and double C-O bonds, which are notoriously
unstable, into the triple valence and stable CO bond.

Stated in different terms, the study of gases created under
extreme magnetic fields establishes that any detection of a
conventional molecular component by infrared or other currently
available spectrometers, by no means, implies that such a
molecular component actually exists in the original gas, because
said molecular component can be created by the instrument itself
at the time of the detection and not be present in the original
gas.

More generally, measurements of magnegases are intriguing inasmuch
as they reminds us that available analytic methods and equipment
were conceived to detect conventional {\sl stable molecules}. As
such, the same methods and equipment are basically insufficient
for true experimental measurements of magnecules beyond the level
of personal beliefs, because magnecules are not as stable as
molecules and they can experience decomposition as well as
mutation into conventional molecules triggered by the available
detecting means themselves, whether the latter are given by
photons or electron beams.

\section*{Acknowledgments}

The authors are grateful to A.V.~Nikolaev, M.I.~Mazhitov,
V.V.~Arkhipov, and A.~Zhumashev for discussions.

\end{document}